\documentclass[
twocolumn,
superscriptaddress,
groupedaddress,
showpacs,
nofootinbib,
nobibnotes,
amsmath,amssymb,
aps,
prl,
]{revtex4}
\usepackage{graphicx}
\usepackage{dcolumn}
\usepackage{bm}

\newcommand{\abs}[1]{\left\vert#1\right\vert}

\begin{document}

\preprint{APS/123-QED}

\title{Fragmentation dynamics of DNA sequence duplications}

\author{M.V. Koroteev and J. Miller}
\email{maxim.koroteev@oist.jp}
 
\affiliation{%
 Physics and Biology Unit, Okinawa Institute of Science and
 Technology (Graduate University)\ Kunigami 1919-1, Onna-son, Okinawa 904-2234,
 Japan }%
\begin{abstract}
Motivated by empirical observations of algebraic duplicated sequence length distributions in
a broad range of natural genomes, we analytically formulate and solve a class of simple discrete 
duplication/substitution models that generate steady-states sharing this property. Continuum equations
are derived for arbitrary time-independent duplication length source distribution, a limit that we show
can be mapped directly onto certain fragmentation models that have been intensively studied by physicists
in recent years. Quantitative agreement with simulation is demonstrated. These models account for
the algebraic form and exponent of naturally occuring duplication length distributions without
the need for fine-tuning of parameters.
\end{abstract}

\maketitle

A century has elapsed since the earliest reports of the evolutionary impact 
of gene duplication\cite{taylor}. At the time there existed only a macroscopic and phenomenological
conception of genetic material, but within the last decade static characterization of the
finest details of the latter has become routine. Its dynamics, on the other
hand, remains for the most part only indirectly accessible; \lq snapshots\rq  of
complete individual genomes at short time intervals are not yet practical, and dynamics must be inferred from their cumulative effect on representative genome sequences.

This dynamics is important because to a good approximation genome sequence determines,
via natural selection, the fates of individuals and of species - but our understanding of how this happens
is primitive. Contemporary lineages are our primary source of genome sequence,
making it difficult to associate the presence or absence of most genomic features with 
their effects, if any, on an individual. Indeed, a primary goal of modern genomics
is to determine if, when, and on what time scales the sequence evolution reflects selection.

{\it Neutral} models of sequence evolution - sequence dynamics that, on the time scales of interest,
are independent of selection - underlie all methods that we know of to achieve
this goal \cite{stone}. When sequences common to two different organisms, or that appear multiply within the same genome,
exhibit identity exceeding (falling short of) that expected on given model
of neutral evolution, it is taken as evidence either that negative (positive) 
selection is acting on these sequences, or that the neutral model is
flawed. As a given sequence fragment has some chance of exhibiting any excess or shortfall of identity within a 
neutral model, selection is inferred probablistically by studying frequencies
of the levels of sequence identity within or between genomes \cite{stone}. Thus,
length distributions of similar or identical sequences have traditionally played a fundamental role in genomics and molecular genetics,
and our interpretation of genomic sequence relies upon our understanding of these distributions.

The topic of this manuscript originates in an empirical observation of algebraic duplicated-sequence
length distributions in a broad range of natural genomes \cite{miller_report,
halvak, gao_miller, taillefer_miller}. In \cite{koroteev_miller2011} an empirical model of duplication
was proposed that accounted for the observed algebraic distribution
of duplicated sequence lengths in natural genomes, but relied on tuning the length distribution
of the duplication source. Here we analytically derive and solve an
alternative model for which no such tuning is required.

The action of duplication is to copy a sequence fragment and subsequently to insert it,
or to substitute it for a same-sized sequence fragment, elsewhere in a
chromosome \cite{graur}. Standard models of sequence evolution also incorporate 
random, uncorrelated base substitution.

A chromosome consists of a string of $L$ bases chosen from a finite alphabet; in natural genomes
the alphabet is typically represented by four bases A, G, C, and T;
for simplicity and without loss of generality we use here a two base alphabet.
The fundamental sequence element that we study is the the set of repeated sequences
within the chromosome, counted in an algorithm-independent way. Specifically, we
study \lq supermaximal repeats\rq\ (or \lq super maxmers\rq): sequence duplications neither copy
of which is contained in any longer sequence duplication within
the chromosome \cite{taillefer_miller, koroteev_miller2011}.
From now on, we refer to a supermaxmer of length $m$ simply as an $m$-mer.

Within our models duplications occur with the rate $\beta$ per unit time: namely, a subsequence of the 
length $m$ is chosen randomly within the chromosome
according to a predetermined source distribution $P(m)$
and is susbtituted for a sequence of length $m$ at another randomly chosen position in the chromosome. 
It was numerically demonstrated \cite{koroteev_miller2011} that for monoscale sources
and certain power-law source distributions, the duplication length distribution attained a stationary
state at long times.

We denote the ensemble-averaged number of $m$-mers at time $t$ as $f(t,m)$. 
For a monoscale source $P(m)=\delta_{c}(D-m)$ ($\delta_{c}$, Kronecker delta),
we expect at stationarity that $f(t,m)$ will decay rapidly for $m > D$. There are two processes altering the number of $m$-mers:
a new duplication of fixed length $D$ can fragment an existing $m$-mer or
generate a new $m$-mers by fragmenting a longer $m$-mer; processes of
higher order in $D/L$ are ignored, where $D, L>>1$ but $D<<L$.

The time dependence of $f$ can be represented by terms of the form $uf(t,m)$,
$u$ being a coefficient describing the rate of creation or destruction of 
corresponding $m$-mers. 
An $m$-mer is annihilated when a newly created duplication of length $D$ overlaps
with one of the sequences composing the $m$-mer; the rate of this event is
$2(D+m-a)/L$, where $a$ is the length of a single
base. Alternatively an $m$-mer may be created when a newly created
duplication overlaps with an $m+k$-mer, $k > 0$. The probability that a
$m+k$-mer produces an $m$-mer is $4a/L$.

Supermaxmers may also be annihilated by base substitution.
Substitution occurs with rate $\mu$ per time unit per unit
length (in bases); duplication with rate $\beta$,
measured in $1/$time unit. Then
the balance equation takes the form
$$
f(t+1,m)- f(t,m)= - 2\left[\frac{m+D-a}{L}\beta+\mu m\right]f(t,m) +
$$
\begin{equation}
\label{pm}
+\left[\frac{4}{L}a\beta + 4a\mu\right]\sum_{k=m+1}^{D}f(t,k) +
2\beta\delta_{c}(D-m). 
\end{equation}

The dimensions of (\ref{pm}) are correct as we take $\Delta t=1$, as it is seen
from lhs of the equation.  The solution of (\ref{pm}) converges to a stationary
one. To see this, take $\mu=0$, $\beta=1$, $a=1$ [base], and note that
(\ref{pm}) may be represented in matrix form as $\vec{f}(t+1)=A\vec{f}(t) + \vec{\delta_{c}}$, where the matrix $A$ is 
such that $A_{ii} = 2(1-(D+1)/L)$, $A_{ij}=4/L$ for
$i<j$, and $A_{ij}=0$, for $i>j$, $i,j=1,2,\ldots D$; the vector
$\vec{\delta}_{c}$ is $\delta_{D}=2$, $\delta_{i}=0$, $i<D$.
The matrix is upper triangular and its eigenvalues are readily computed
yielding $\lambda_{i} = 1 - \beta(D+ (i-1))/L$. It
is evident that $0<\abs{\lambda_{i}}<1$ for all $i$, as we assumed
$D<<L$ and $i=1,2,\ldots D$, and consequently, the iteration is
guaranteed to converge. For $\mu\ne 0$ the eigenvalues have the form
$\lambda_{i}=1-\beta\frac{i+D-1}{L}-\mu i$, thus the
requirement of the convergence to a stationary state $\abs{\lambda}<1$ yields
(approximately) $D<<L$, $\mu\Delta\tau <1/D$, $\Delta\tau$ being a time step. 

If some initial state $\vec{f}(0)$ is given and if we denote by $\vec{f}_{s}$ the limiting
stationary state of the system, we can calculate $\vec{f}_{s}$ as follows  
$$
\vec{f}_{s}=\lim_{t\to\infty}\vec{f}(t) = 
\lim_{t\to\infty}\left[T\sum_{k=0}^{t-1}\Lambda^{k}T^{-1}\vec{\delta} +
T\Lambda^{t}T^{-1}\vec{f}(0)\right]
$$
\begin{equation}
\label{limit-stationary}
= 
T\lim_{t\to\infty}\sum_{k=0}^{t-1}\Lambda^{k}T^{-1}\vec{\delta}, 
\end{equation}
where $A=T\Lambda T^{-1}$ and $T$ diagonalizes $A$ and consists of eigen vectors
of $A$. Further estimates show that 
$\vec{f}_{s} = LT\tilde{\Lambda}T^{-1}\vec{\delta}$, where, e.g., for the case
$\mu=0$ we have $\tilde{\Lambda}_{ji}= 1/(D+i-1))$, for $j=i$ and
$\tilde{\Lambda}_{ji}=0$ when $j\neq i$. We may write down the exact
stationary solution of the equation (\ref{pm}) in scalar form 
$$
f(m,D,L,\mu) =
\left[ \frac{D-(m-a)}{\beta\frac{D+(m-a)}{L} +m\mu}
-2\frac{D-m}{\beta\frac{D+m}{L} +(m+a)\mu}\right.
$$ 
\begin{equation} 
\label{pm.equ}
\left.
+ \frac{D-(m+a)}{\beta\frac{D+(m+a)}{L}
+(m+2a)\mu}\right], \quad m<D.
\end{equation}

\begin{figure}
\includegraphics[width=240pt, height=160pt]{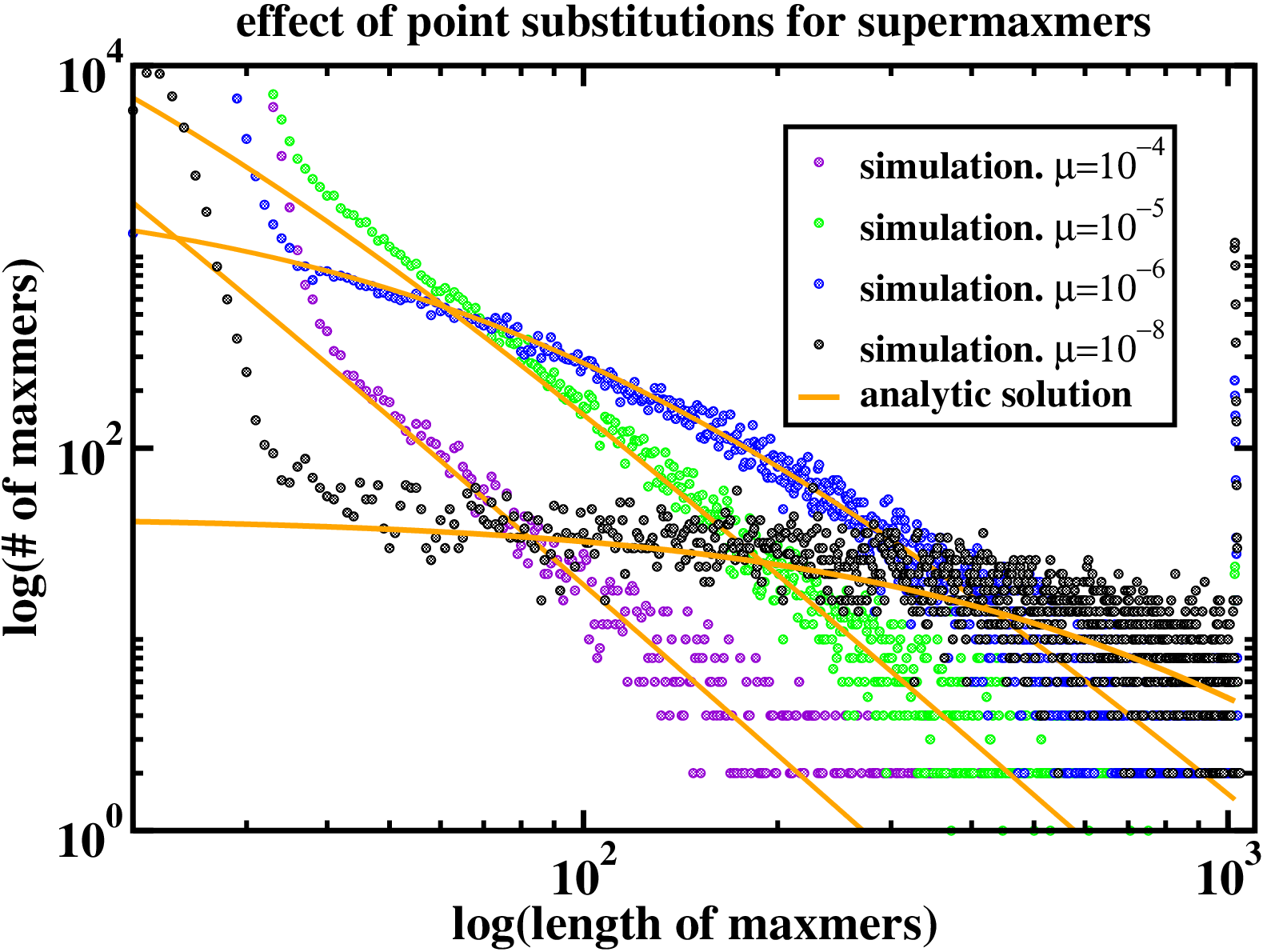}   
\caption{Curves represent stationary states of the system described in
the paper for various base substitution rates
$\mu$ and corresponding analytic solutions of the system (\ref{pm}) for $\beta=1$. The
chromosome length $L=10^{7}$; source length $D=1024$. Increasing base substitution rates
exhibits a power-law tail with the exponent $-3$. Note also the match
of amplitudes between simulations and solution.}
\label{fig:pmfig}
\end{figure}

Obvious scaling wrt. $L$ is observed when $\mu=0$.
Comparisons to the empirical model \cite{koroteev_miller2011} with
(\ref{pm.equ}) for $\mu\ne 0$ are presented in fig. \ref{fig:pmfig}.
It is evident that with increasing base substitution rate both simulations and
the solution demonstrate power-law behavior with the exponent $-3$. To obtain this
exponent from the solution (\ref{pm.equ}) observe that $f$ in (\ref{pm.equ})
is approximately represented as $g^{\prime\prime}(x)$, where
$g(x)=(D-x)/(\beta\frac{D+x}{L} +\mu x)$. Then 
one obtains $f(m, D, L, \mu)\sim 1/(\beta\frac{D+m}{L} + m\mu)^{3}$.
The peak observed in the left-hand side of the length distributions reflects
that of a random sequence of the length $L$. High mutations
conserve this random part of the distribution while duplications tend to
distort the statistic. The maximum of the peak is located in the point
$m=\log_{2}L$ for binary sequence. For the maximum length of supermaxmers in a
random binary sequence there is an estimate $M_{L}\sim 2\log_{2}L$\cite{arratia}
which thus corresponds to the width of the peak.

For dynamics described by a power-law source
$p(m)\sim 1/m^{\gamma}$ the characteristic scale $D$ is replaced by the first moment of the source.
We also make all lengths dimensionless, dividing them by the length of
$1$ base $a$ or by the first moment of the source $M_{1}$.
Then, we have for probabilities of fragmentation $p(m)
=1/(m^{\gamma}\phi_{1}(\gamma,N))$, where $\phi_{1}(\gamma,N)=
\zeta(\gamma)-\zeta(\gamma, N+1)$, $L=Na$, and $\zeta(\gamma)$ is the Riemann zeta-function, $\zeta(\gamma, N+1)$ is the
generalized zeta-function. The equation for a finite-size system with a power-law source can be obtained
similarly to that for the monoscale source and has the form
$$
\Delta f(t,m)= - 2\left[
\frac{1}{N}\sum_{r=1}^{N}\frac{m+r-1}{\phi_{1}(\gamma,N)r^{\gamma}}\beta +
ma\mu\right]f(t,m)+
$$
\begin{equation}
\label{equation-power-law}
 + \left[ \frac{4}{N}\beta + 4a\mu \right]\sum_{k=m+1}^{N}f(t,k) +
2\beta\frac{1}{\phi_{1}(\gamma,N)m^{\gamma}},
\end{equation}
and $m,r=1,2,\ldots, N$.

\begin{figure}
\includegraphics[width=240pt, height=160pt]{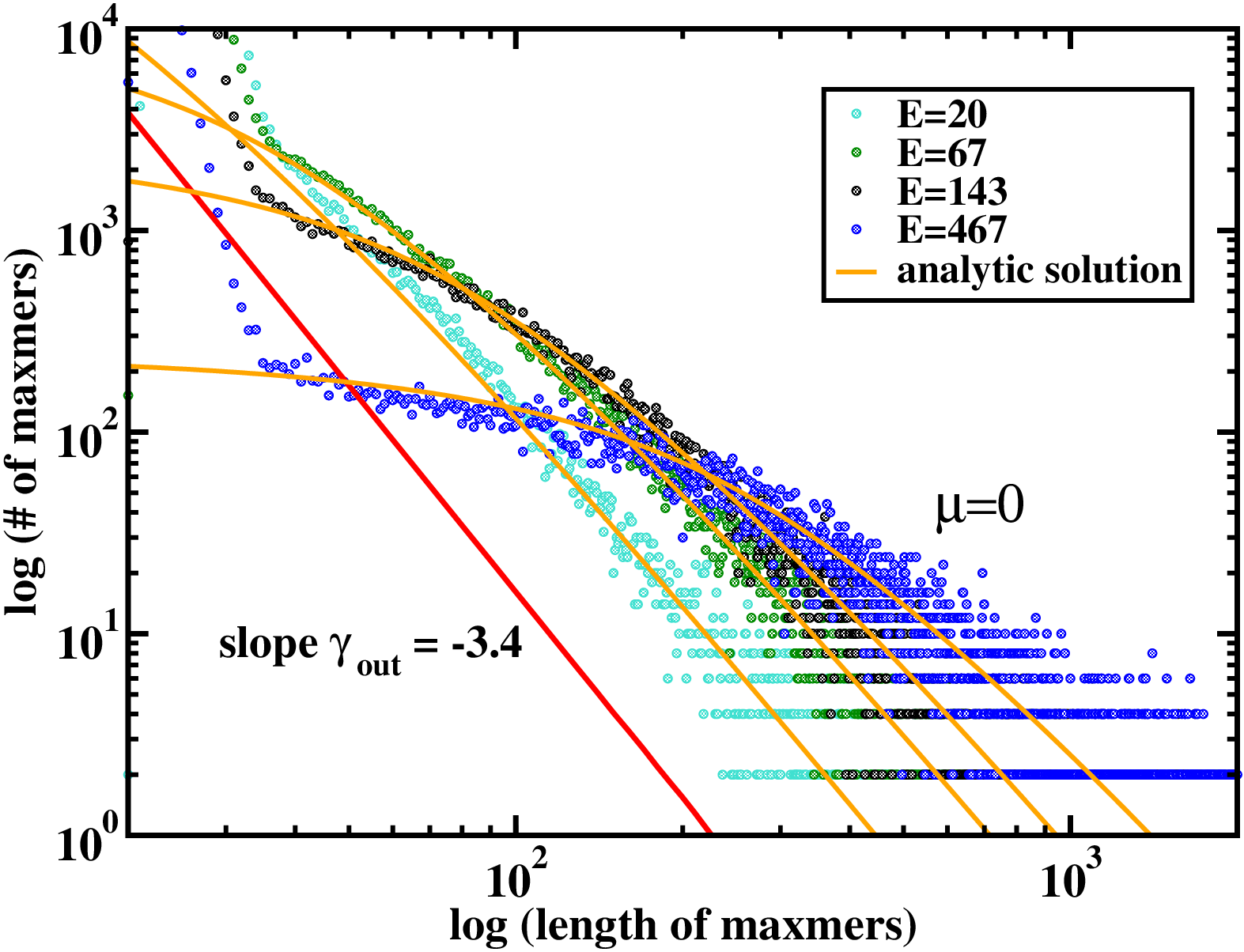}   
\caption{Curves represent stationary states of the system described in
\cite{koroteev_miller2011} for various first moments $M_{1}$ of the power-law source
of the form $p(\bar{x})\sim 1/(\bar{m}_{0}+\bar{x}^{\gamma})$ compared to the
solution of (\ref{equation-power-law}). $L=10^{7}$,
$\mu=0$, $\gamma=-2.4$. Deviation of solution from simulations is observed for small $M_{1}$, when the condition $M_{1}>>a$ is violated. 
For large lengths it 
is observed the regime with the exponent
$\gamma+1$ described in \cite{koroteev_miller2011}. Values $\bar{m}_{0}$
corresponding to $M_{1}$ on the plot are: 10, 35, 75, 250.}.
\label{fig:power-law-figure}
\end{figure}

The structure of the source makes an analytic representation unwieldy; the 
solution of (\ref{equation-power-law}) as $t\to\infty$ was
obtained numerically. The comparison of the solution with simulations
is presented in fig. \ref{fig:power-law-figure}.
Evidently equation (\ref{equation-power-law}) reproduces both
exponent and amplitude and thus the dynamics involved encompasses both small and
large mutation rates.
\begin{figure}
\includegraphics[width=240pt, height=160pt]{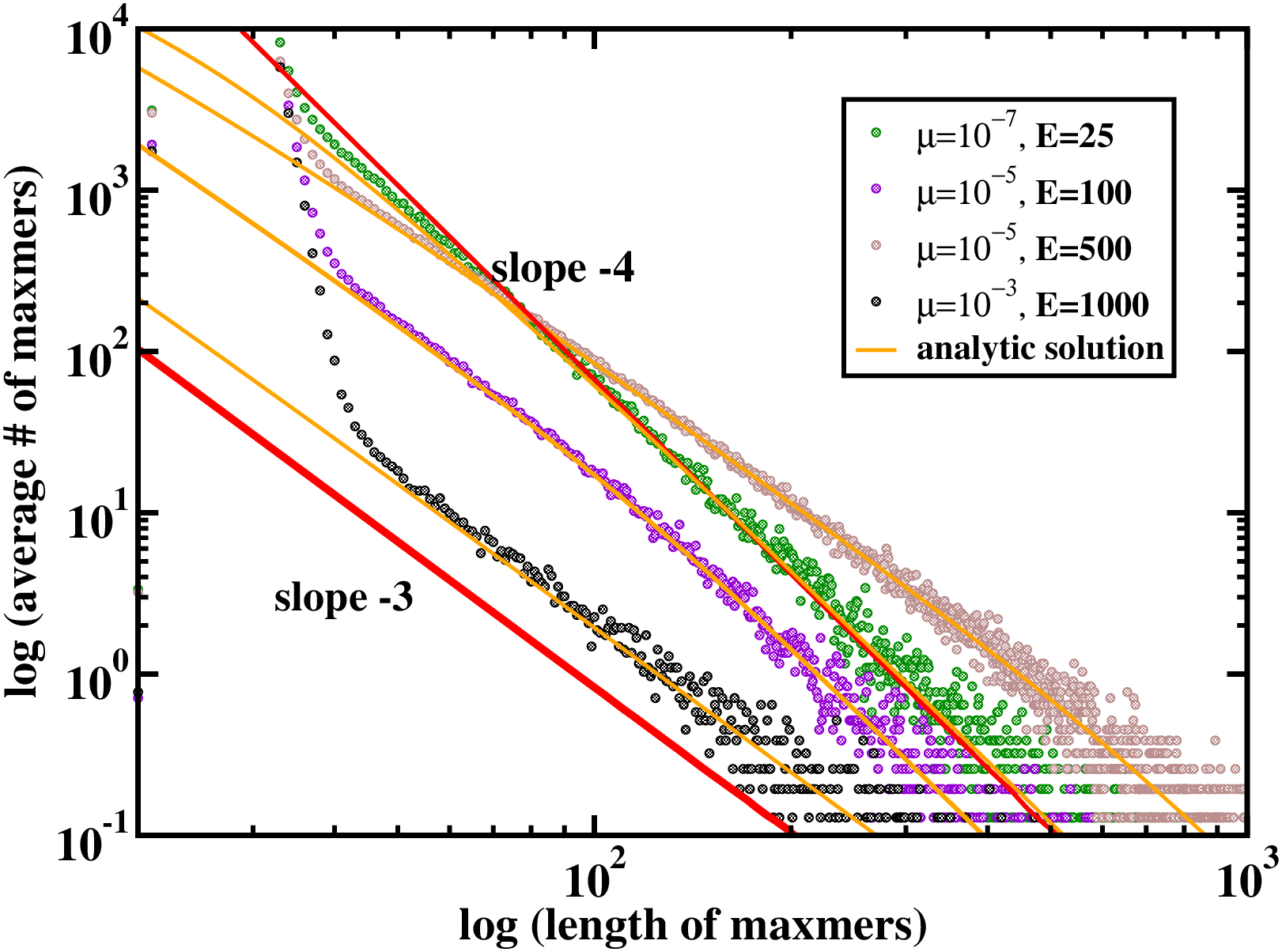}    
\caption{Curves represent stationary states of the system described in
the Letter for $L=10^{7}$ and varying $M_{1}$ and $\mu$ compared to the solution
of (\ref{equation-power-law}). The distributions are obtained by averaging
over $100$ realizations. These are shown the regime $\mu^{-1}\approx M_{1}$
corresponding to the continuum equation and the regime with small 
mutation rate $\mu$ studied in \cite{koroteev_miller2011}, $\gamma=-3.0$.}.
\label{fig:power-law_mutations}
\end{figure}

Finally, some continuum limits following from these discrete
dynamics are obtained. As all lengths are measured in bases we dimensionalize them as
follows: $\bar{a}=a/M_{1}$, $\bar{m}=m/M_{1}$, $\bar{L}=L/M_{1}$,
$\bar{t}=t\beta$, $\bar{\mu}=M_{1}\mu/\beta$. Then (\ref{equation-power-law}) with
an arbitrary source $P(\bar{m})$ takes the form
$$
\frac{\Delta f(t,m)}{\Delta t}= - 2\left[
\frac{1+\bar{m}-\bar{a}}{\bar{L}} +
\bar{\mu}\bar{m}\right]f+
$$
\begin{equation}
\label{equation-power-law2}
 + \left[ \frac{4\bar{a}}{\bar{L}} + 4\bar{a}\bar{\mu}
 \right]\sum_{k=m+1}^{N}f + 2P(\bar{m}).
\end{equation}
Taking $\bar{a}\to 0$ and
introducing densities $f=\bar{a}\hat{f}$, $P=\bar{a}\hat{p}$ with
$\bar{L}=N\bar{a}$, $\bar{m}=n\bar{a}$, the equation becomes
$$
\frac{\Delta\hat{f}(t,m)}{\Delta t}= - 2\left[
\frac{1+n\bar{a}-\bar{a}}{N\bar{a}} +
\bar{\mu}\bar{a}n\right]\hat{f}+
$$
\begin{equation}
\label{equation-power-law3}
 + \left[ \frac{4}{N\bar{a}} + 4\bar{\mu}
 \right]\sum_{k=m+1}^{N}\hat{f}\bar{a} + 2\hat{p}(\bar{m}).
\end{equation}
Additionally, we introduce a duplication rate $\lambda$ measured per base taking
$\beta=\lambda L$.

As $\bar{a}\to 0$ we assume $a<<M_{1}$, and as $\mu$ and $\lambda$ diminish with
$a$, we need to evaluate the orders of corresponding terms. We keep the
product $n\bar{a}$ finite and denote it by $x$; this implies $n\sim\bar{a}^{-1}$
and corresponds to an intermediate regime\cite{barenblatt} for (\ref{equation-power-law2}). The source term
has the order $\sim 1$. The main duplication term in
(\ref{equation-power-law3}) has the order $\sim\bar{a}^{\alpha-1}$, and
mutation terms the order $\bar{\mu}\sim (\mu/\lambda)\bar{a}^{\alpha-1}$ as
$\bar{a}\to 0$. We consider the case $\bar{\mu}\sim 1$,
$\bar{a}^{\alpha-1}=o(1), \bar{a}\to 0$, other regimes being described
elsewhere. Taking into account that in this regime $\bar{L}\to\infty$ and $L>>M_{1}$, the equation takes the form
\begin{equation}
\label{continuum_equation}
\frac{\partial\hat{f}}{\partial\bar{t}}= -2x\bar{\mu}\hat{f} +
4\bar{\mu}\int_{x}^{\infty}\hat{f}(\bar{t},y)dy + 2\hat{p}(x) +
O(\bar{a}^{\alpha-1}).
\end{equation}
The main order regime corresponds to fragmentation with input
studied in \cite{ben-naim}. 

For the numerical simulations in fig. \ref{fig:pmfig} we set $L=10^{7}$,
$M_{1}=D\approx 10^{3}$ and vary $\mu$; thus, $L>>M_{1}$. As $\mu$ approaches
$10^{3}$ the output distribution approaches an algebraic form with exponent
$-3$, corresponding to the solution of the fragmentation equation for
a monodisperse source\cite{ben-naim, krapivsky}. Fig. \ref{fig:power-law-figure}
corresponds to the regime with vanishing $\mu$ and is not described by
(\ref{continuum_equation}); fig. \ref{fig:power-law_mutations}
demonstrates various regimes and exponent $-3$ which is observed for $M_{1}>>a$,
$\bar{a}^{\alpha-1}=o(\bar{\mu})$ and $m<<M_{1}$.
 
\begin{figure} 
\includegraphics[width=240pt, height=160pt]{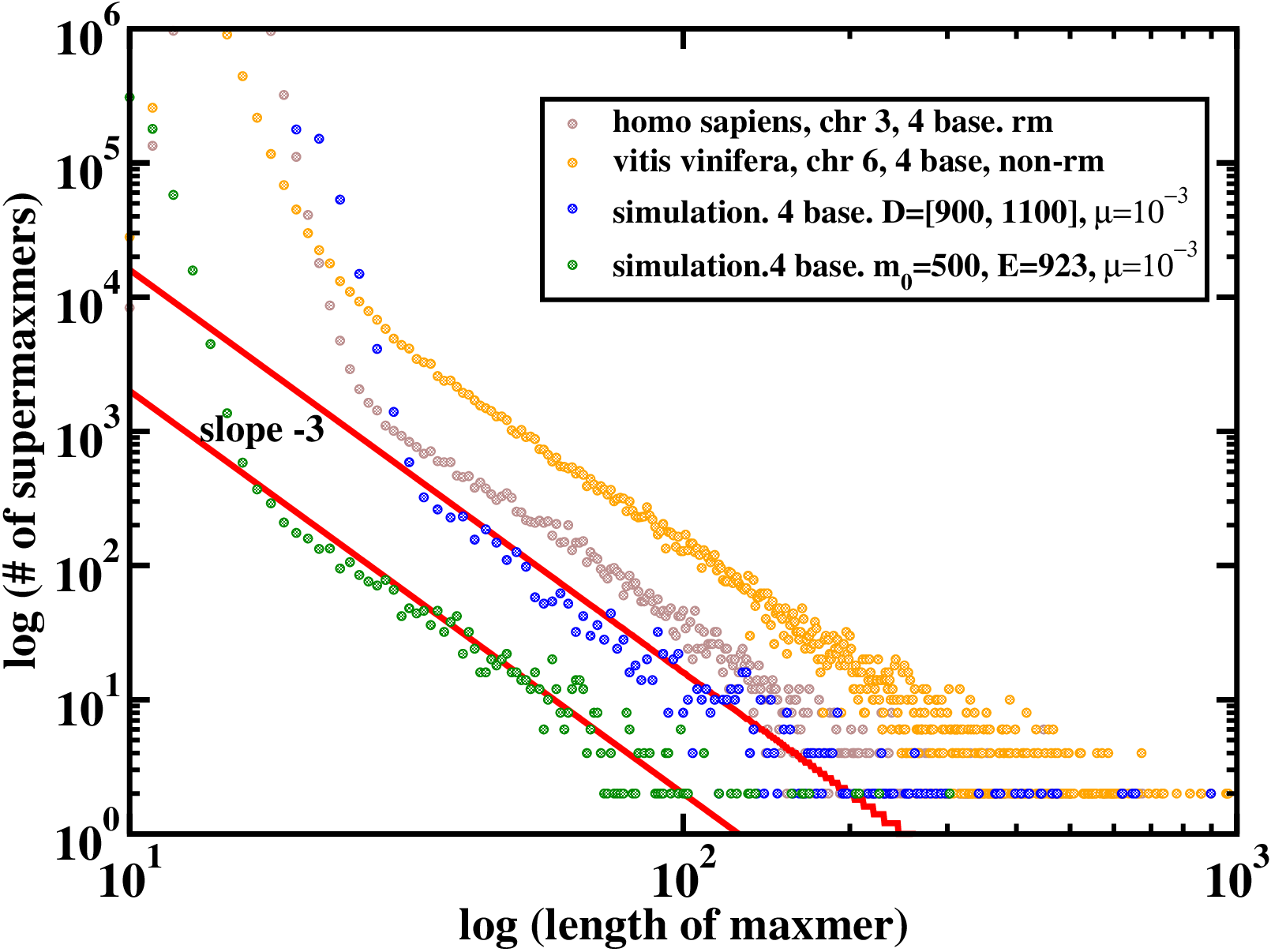}     
\caption{Human chr. 3 (brown) was repeat-masked, grapevine (orange) was not.
Source length distributions were chosen as power-law with exponent 
$-2.4$ (green); and uniform on the interval $[900, 1100]$ (blue). The latter
was shifted along $x$ axis to the right by the factor of $2$ for clarity. Straight lines with the slope $-3$ have the amplitudes
$2M_{1}\beta/\mu$ in accordance with (\ref{estimate}) and the exact solution of
(\ref{continuum_equation}). For the human chr. 3 we have
$2M_{1}\beta/\mu\approx 3\times 10^{7}$, and for vitis vinifera chr. 6, $\approx
10^{8}$.}
 \label{fig:real_genomes} 
\end{figure}

In fig. \ref{fig:real_genomes} we provide comparison of natural data with our
simulations. Both chromosomes, one from human and the other from grapevine, demonstrate good fit to $-3$
that is also reproduced by our models.
 
From (\ref{equation-power-law3}) and (\ref{continuum_equation}) it also follows
that the duplications in this regime are dilute: a duplicated sequence is broken down by
substitutions long before there there is any opportunity for a subsequent
duplication to overlap with it.
In this sense each $m$-mer evolves independently of other $m$-mers.
Therefore, following Ben-Naim \cite{ben-naim} we can estimate the fragment
length distribution by following a typical duplication of length $M_{1}$.
Substitutions break the duplication into fragments whose number $\mathcal{M}$
varies in time as $\mathcal{M}=M_{1}\mu t$ so that the average fragment length
at time $t$ is $\langle m\rangle=M_{1}/\mathcal{M}$. Consecutive mutations are
independent, hence the distribution of fragment lengths is close to Poisson for $\bar{a}<<1$, i.e.,
$p(m)\sim m^{-1}\exp(-m/\langle m\rangle)$, neglecting
contributions exponentially in $M_{1}$. Then the number $S(t,m)$ of fragments
of length $m$ is $\mathcal{M}p(m)$. New duplications occur continuously, so to
obtain the fragment length distribution as $t\to\infty$ we integrate over time to obtain
\begin{equation}
\label{estimate}
S(m)=\int_{0}^{\infty}\mathcal{M}\beta p(m)dt
\sim\int_{0}^{\infty}\frac{\mathcal{M}\beta}{m}e^{-m/\langle m\rangle}dt =
\frac{M_{1}\beta}{\mu m^{3}}.
\end{equation}
The similar result can be also obtained from the exact solution of
(\ref{continuum_equation}) to give $S(m)=2M_{1}\beta/(\mu m^{3})$; the
additional prefactor $2$ appears as in the equations for each sequences we
count another one, which is identical to the former. The consistency of
dimensions follows from the presence of additional prefactor $a$, which is equal
to $1$ for the discrete case.

The estimate allows to
calculate $M_{1}$ from the empirical distributions, e.g., for fig. \ref{fig:pmfig} $\beta=1$, $\mu=10^{-4}$ in the algebraic regime. 
Then we can estimate $S(m)$ for $m=1$ from the plot
\cite{supplemental}(supplemental figure 1)); we have $S(m)\approx 2\times10^{7}$
and from (\ref{estimate}) we find $M_{1}\approx 10^{3}$ which pretty well corresponds to the real value $M_{1}=1024$ for this simulation. Similar estimates can be obtained for different sources.

For eukaryotes, gene duplicatons are conventionally estimated to arise
at around $\sim 10^{-2}$ per gene per $10^{6}y$ (years)\cite{lynch}; assuming
$\sim 10^{4}$ genes per eukaryotic genome yields a genome-wide gene duplication rate\cite{lynch}
$\beta_{0}=10^{2}/10^{6}y$. Thus for the human genome with around
$4\times 10^{4}$ genes, $\beta_{0}\approx
300/10^{6}y$\cite{lynch}. Then, for duplication rate per base we have
$\lambda_{0}=\beta_{0}/L_{0}$, where $L_{0}$ the number of bases belonging to
genes; accepting the estimate $L_{0}$ to be $2\%$ of $L$ we can estimate
$\lambda_{0}$. Let us assume that duplications, as well as point substitutions,
are uniform over genome, i.e., duplication rate per base $\lambda=\lambda_{0}$
and for $\beta$ we have $\beta\approx 50\beta_{0}$. The
average time between duplications, $1/\beta = 200y$ corresponds to a single
time unit in the parametrization of our models. The point substitution rate
is $\mu \sim 10^{-2}$ per base per $10^{6}y$\cite{lynch} or $\sim 10^{-5}$ per 
base per time unit. The algebraic regime with exponent $-3$ is already attained
for $\mu=10^{-5}$ in fig. \ref{fig:pmfig} and $\mu/\lambda=100 >> 1$ for this
simulation.

To give estimates of time of emerging of currently observed
identical repeats in human genome we use repeat-masked chr. 3
(Supplemental figure 2). With estimates and assumptions given above we find
$\lambda\approx 0.5\times10^{-5}$ per base, per $10^{6}$ y and
consequently $\beta\approx 0.5\times10^{3}$ (we take into account here that the
length of repeat-masked chromosome differs significantly from that of the whole
sequence). Then, we find $M_{1}\approx 300$, the average duplication length in human
chr. 3. Note also, that the tail drops off at this length
\cite{supplemental}(supplemental figure 2). Thus $M_{1}\beta\approx 1.5\times
10^{5}$ bases related to supermaxmers were duplicated in human genome per $10^{6}$ years. The region of the tail in supp. fig. 2 corresponds to
lengths $m>30$; the tail contains $\approx 10^{6}$ bases, hence, assuming
similar processes in different chromosomes, the observed long identical
duplicates occurred in the human genome last $6-7$ million years. This estimate
fairly well corresponds to divergence time between human and chimp.

A $-3$ exponent of distributions is observed in many
natural genomes\cite{taillefer_miller}. It is important to stress that this regime is
reproduced by \cite{koroteev_miller2011} as well as by the model
suggested here, which incorporates \cite{koroteev_miller2011}
as a specific case. 
This regime is, in part, detected when stationary solutions of the
equations (\ref{pm}) and (\ref{equation-power-law}) become weakly dependent on a
duplication source, demonstrating at the same time algebraic form with the
slope $-3$. Thus the state of many currently studied genomes mapped to this
regime of our dynamics, may be understood as a result of continuous interaction
of point substitutions and (segmental) duplications generated by {\it some} source. These results 
also provide some evidence for the neutral nature of long segmental duplications. On the other hand, the
assumptions for (\ref{equation-power-law3}) may be altered to obtain different regimes to include
genomes, whose state deviates from $-3$ regime, e.g.,
because of extensive recent duplications.

As our manuscript was in final stages of revision, we learned
of \cite{ardnt} where a similar dynamics to that studied here was introduced.

Compared to \cite{ardnt} we derive continuum dynamics directly and show how the
crucial parameter $M_{1}$ (or $D$), the first moment of the source, appears in
 calculations to determine the regime in which we observe genomes with the exponent $-3$ for the length
distribution. 

In addition, our dynamics
treats a broader problem in two respects: 1) we introduce and demonstrate the
dependence of the observed regime in length distributions on $M_{1}$ as there is only one specific
regime with the exponent $-3$ in which the first moment $M_{1}$ turns out to be less
important; 2) our dynamics allows to consider various asymptotic orders of
$\mu/\lambda$, not necessarily $\mu>>\lambda$.

Thus, $-3$ is observed in
(\ref{continuum_equation}) asymptotically as $x\to 0$, corresponding to $m<<M_{1}$ for discrete equations (\ref{pm}), 
(\ref{equation-power-law}), i.e.,
the algebraic regime with the exponent $-3$
in natural genomes may be observed for duplicate lengths $ << M_{1}$. If the source
produces short duplicates, which are in the same time are not hit by strong
mutations (e.g., $\mu\sim\lambda$) then the tail occurs for $m>>M_{1}$ or in
asymptotic regime $x\to\infty$ for (\ref{continuum_equation}) and we may observe regimes with
different exponents (fig. \ref{fig:power-law_mutations}) or even non-algebraic regimes. The latter ones are also observed in real genomes
\cite{supplemental}(supplemental figure 3) and can not be treated in terms of
specific $-3$ exponent but are reproduced by our dynamics. Thus we can map various asymptotic
regions of parameters to natural sequences to fit our observations in
genomes, demonstrating variety of length distributions for duplicates.

We acknowledge Kun Gao, Eddy Taillefer and Satish Venkatesan for helpful
duscussion of this work. We thank Quoc-Viet Ha for the help with computations.
We also thank Peter Arndt, Florian Massip, Nick Barton, Daniel Weissman, and
Tiago Paixao for discussions of this work and the manuscript.





\end{document}